\documentclass[twocolumn,showpacs,superscriptaddress,amsmath,amssymb]{revtex4}
\usepackage{graphicx}
\usepackage{dcolumn}
\usepackage{bm}
\usepackage{booktabs}
\usepackage{color}
\usepackage{mathrsfs}
\usepackage{ulem}
\usepackage{slashbox}

\begin{document}
\title{ High cumulants from the 3-dimensional $O(1, 2, 4)$ spin models}

\author{Xue Pan} 
\affiliation{Key Laboratory of Quark and Lepton Physics (MOE) and
Institute of Particle Physics, Central China Normal University, Wuhan 430079, China}
\author{Lizhu Chen} 
\affiliation{Key Laboratory of Quark and Lepton Physics (MOE) and
Institute of Particle Physics, Central China Normal University, Wuhan 430079, China}
\author{X.S. Chen} 
\affiliation{Institute of Theoretical Physics, Chinese Academy of Sciences,
        Beijing 100190, China  }
\author{Yuanfang Wu} 
\affiliation{Key Laboratory of Quark and Lepton Physics (MOE) and
Institute of Particle Physics, Central China Normal University, Wuhan 430079, China}

\begin{abstract}

Considering different universality classes of the QCD phase transitions, we perform the Monte Carlo simulations of the 3-dimensional $O(1, 2, 4)$ models at vanishing and non-vanishing external field, respectively. Interesting high cumulants of the order parameter and energy from $O(1)$ (Ising) spin model, and the cumulants of the energy from $O(2)$ and $O(4)$ spin models are presented. The generic features of the cumulants are discussed. They are instructive to the high cumulants of the net baryon number in the QCD phase transitions.

\end{abstract}

\pacs{25.75.Nq, 21.65.Qr, 75.10.HK}

\maketitle
\section{Introduction}

A thermodynamic system may undergo a phase transition when temperature and/or pressure vary. For the strong interacting QCD system, it is known that the hadronic matter will change to partonic phase at very high temperature and density~\cite{QGP-th}. The data from ultra-relativistic nuclear-nuclear collisions at RHIC has shown that the partonic phase of QCD matter--quark-gluon plasma (QGP) has been formed~\cite{QGP-ep}. One of the main goals of current ultra-relativistic heavy-ion
collision experiments is to map the QCD phase diagram~\cite{Star-BES}.

The critical point at the QCD phase boundary is particular interested, as the large fluctuations are expected. The sensitive probes of QCD critical point are the high cumulants of the conserved charges, i.e., net baryon number, net electric charge, and net strangeness~\cite{stephanov-prl91, koch, Stephanov-prl102, Asakawa-prl103, Staphanov-prl107, six order}.

The net baryon number fluctuations have been studied using lattice QCD simulations and QCD effective models~\cite{M. Cheng, Gupta-plb, Fu weijie-PNJL, PQM model}. However, due to the difficulties of the lattice calculations and model estimations, the high cumulants of the net baryon number are still not final~\cite{Karsch-PRD84,Review-Budpast}.
The universality of the critical fluctuations allows us to study the relevant cumulants in a simple system, where the high cumulants can be precisely obtained.

It is known that the QCD critical point, which terminates the first order phase transition line, is in the same universality class as the 3-dimensional Ising model~\cite{class 1,class 2, class 3,class 4}. So the high cumulants of order parameter and energy in the vicinity of the critical point of the 3-dimensional Ising model are called for. They should be a good reference for various relevant calculations.

In the chiral limit, the chiral phase transition for 2-flavor QCD belongs to the same universality class as the 3-dimensional $O(4)$ spin model~\cite{class 4}. The singular behavior of the net baryon number fluctuations is expected to be governed by the universal $O(4)$ symmetry group~\cite{symmetry group}. Owing to the staggered fermions in lattice calculations, the (2+1)-flavor chiral phase transition may fall into the 3-dimensional $O(2)$ universality class~\cite{O2 class 1,O2 class 2,O2 class 3}.

In this paper, we present the high cumulants of the order parameter and energy in the vicinity of the critical temperature from the 3-dimensional Ising model, and energy from the 3-dimensional $O(2)$ and $O(4)$ spin models, respectively. The simulation is performed in a finite-size system. The obtained results should be instructive to a finite system formed in ultra-relativistic heavy-ion collisions.

The paper is organized as follows, firstly, the cumulants of the order parameter and energy from the $O(N)$ spin models are derived in section II. Then, their relations to the net baryon number fluctuations in the QCD phase transitions are discussed. In section III, the high cumulants from the 3-dimensional Ising, $O(2)$ and $O(4)$ spin models are presented and discussed. Finally, the generic features of the cumulants from the three models are summarized in section IV.

\section{Cumulants in the $O(N)$ spin models}

The $O(N)$-invariant nonlinear $\sigma$-models ($O(N)$ spin models) are defined as,
\begin{equation}\label{Hamiltonian}
\beta \mathcal{H} =-J\sum_{\langle i,j\rangle} \vec{S}_{i}\cdot \vec{S}_{j}-\vec H\cdot\sum_{i} \vec{S}_{i},
\end{equation}
where $\mathcal{H}$ is the Hamiltonian, $J$ is an interaction energy between nearest-neighbor spins $\langle i,j\rangle$, and $\vec H$ is the external magnetic field. $J$ and $\vec H$ are both reduced quantities which already contain a factor $\beta=1/T$. $\vec S_{i}$ is a unit vector of $N$-components at site $i$ of a $d$-dimensional hyper-cubic lattice. It is usually decomposed into the longitudinal (parallel to the magnetic field $\vec H$) and the transverse component
\begin{equation}\label{spin}
\vec S_i=S_i^\parallel \vec e_H+\vec S_i^\perp,
\end{equation}
where $\vec e_H=\vec H/H$. For the 3-dimensional Ising, $O(2)$, and $O(4)$ spin models, $d$ = 3, and $N$ = 1, 2 and 4, respectively.

The (reduced) free energy per unit volume is
\begin{equation}\label{free energy density}
f(T,H) = -\frac{1}V \ln Z,
\end{equation}
where $Z=\int\prod_i d^NS_i \delta(\vec S_i^2-1) \exp(-\beta E+HVS^\parallel)$ is the partition function. $E=-\sum_{\langle i,j\rangle} \vec{S}_{i}\cdot \vec{S}_{j}$ is the energy of a spin configuration, $S^\parallel=\frac{1}{V}\sum_{i}S_{i}^\parallel$ is the lattice average of the longitudinal spin components, $V=L^3$  is the volume, and $L$ is the number of lattice points of each direction.

As we known, the cumulants of the order parameter are the derivatives of the free energy density (Eq.~\eqref{free energy density}) with respect to $H$. We can get the cumulants of the order parameter from the generating function~\cite{generating function},
\begin{equation}\label{the generating function 1}
\left.\kappa_n^S=\frac{d^n}{dx^n}\ln\langle e^{xS^\parallel}\rangle\right|_{x=0}.
\end{equation}
So the first, second, third, fourth and sixth order cumulants of the order parameter are as follows,
\begin{equation}\label{cumulants of order parameter}
\begin{split}
&\kappa_1^S=\langle S^\parallel\rangle,\quad  \kappa_2^S={\langle \delta {S^\parallel}^2 \rangle},\quad  \kappa_3^S={\langle \delta {S^\parallel}^3 \rangle},\\& \kappa_4^S=\langle \delta {S^\parallel}^4 \rangle-3\langle \delta {S^\parallel}^2 \rangle^2, \\&
\kappa_6^S=\langle \delta {S^\parallel}^6 \rangle-10\langle \delta {S^\parallel}^3 \rangle^2+30\langle \delta {S^\parallel}^2 \rangle^3\\& \qquad -15\langle \delta {S^\parallel}^4\rangle\langle \delta {S^\parallel}^2\rangle,
\end{split}
\end{equation}
where $\delta {S^\parallel}=S^\parallel-\langle S^\parallel\rangle$, and $\kappa_1^S$ is the magnetization
(order parameter) of the system. At vanishing external magnetic field, due to the spatial rotation symmetry of the $O(N)$ groups, such defined order parameter is zero. In the case, an approximated order parameter definition is suggested as, $M=\langle|\frac{1}{V}\sum_i{\vec S_i}|\rangle$~\cite{order parameter}.

On the other hand, the cumulants of the energy are the derivatives of the free energy density with respect to the temperature $T$. The generating function is
\begin{equation}\label{the generating function 2}
\left.\kappa_n^E=\frac{d^n}{dx^n}\ln\langle e^{xE}\rangle\right|_{x=0}.
\end{equation}
So the first, second, third, fourth and sixth order cumulants of the energy are as follows,
\begin{equation}\label{first order derivatives of free energy to t}
\begin{split}
&\kappa_1^E={\langle E\rangle},\quad \kappa_2^E={\langle \delta E^2 \rangle},\quad \kappa_3^E={\langle \delta E^3 \rangle},
\\& \kappa_4^E={\langle \delta E^4 \rangle-3\langle \delta E^2 \rangle^2},
\\& \kappa_6^E=\langle \delta E^6 \rangle-10\langle \delta E^3 \rangle^2+30\langle \delta E^2 \rangle^3\\& \qquad -15\langle\delta E^4 \rangle\langle \delta E^2 \rangle,
\end{split}
\end{equation}
where $\delta E=E-\langle E\rangle$.

In the vicinity of the critical point, the free energy density (Eq.(\ref{free energy density})) can be decomposed into two parts, the regular and singular parts. The critical related fluctuations are determined by the singular part. It has the scaling form
\begin{equation}\label{scaling free energy}
f_s(t,h)=l^{-d}f_s(l^{y_t}t,l^{y_h}h).
\end{equation}
Here $t={(T-T_{c})}/T_0$ and $h=H/H_0$ are reduced temperature and magnetic field, $T_0$ and $H_0$ are the normalized parameters. $T_{c}$ is the critical temperature. $y_t$ and $y_h$ are universal critical exponents. In our simulation, we set $J=\beta$ and choose the approximate critical temperatures, $T_{c}^{\rm Ising} =4.51$~\cite{order parameter}, $T_{c}^{O(2)} =2.202$~\cite{O24Tc}, and $T_{c}^{O(4)} =1.068$~\cite{O24Tc} for the 3-dimensional Ising, $O(2)$ and $O(4)$ spin models, respectively.

In order to map the result of the 3-dimensional Ising model to that of the QCD, the following linear ansatz is suggested~\cite{linear map 1,linear map 2,linear map 3},
\begin{equation}\label{linear mapping}
t\approx T-T_{cp}+a(\mu-\mu_{cp}),~ h\approx\mu-\mu_{cp}+b(T-T_{cp}).
\end{equation}
$T_{cp}$ and $\mu_{cp}$ are the temperature and chemical potential at the QCD critical point, respectively. $a$ and $b$ are two parameters determined by QCD. The baryon-baryon correlation length diverges with the exponent $y_t$ and exponent $y_h$ when the critical point is approached along the $t$-direction and $h$-direction, respectively~\cite{Gupta-plb}.

The cumulants of the net baryon number are the derivatives of the QCD free energy density with respect to $\mu$. In the vicinity of the critical point, it is the combination of the derivatives with respect to $t$ and $h$ in the 3-dimensional Ising model. Since $y_h$ is larger than $y_t$~\cite{Ising exponents}, the critical behavior of the net baryon number fluctuations is mainly controlled by the derivatives with respect to $h$, i.e., the fluctuations of the order parameter in the 3-dimensional Ising model.

The singular part of the free energy density for the chiral phase transition is suggested as~\cite{six order}
\begin{equation}\label{QCD free energy}
\frac{f_s(T,\mu_q,h)}{T^4}=Ah^{(1+1/\delta)}f_f(z),~ z=t/h^{\beta\delta},
\end{equation}
where $\beta$ and $\delta$ are the universal critical exponents of the 3-dimensional $O(4)$ spin model. $f_f(z)$ is the scaling function. The reduced temperature $t$ and external field $h$ are expressed as follows
\begin{equation}\label{parameter}
t\equiv\frac{1}{t_0}(\frac{T-T_c}{T_c}+\kappa_\mu(\frac{\mu_{q}}{T})^{2}),~ h\equiv\frac{1}{h_0}\frac{m_q}{T_c}.
\end{equation}
Here $T_c$ is the critical temperature in the chiral limit. $\kappa_{\mu}$ is a parameter determined by QCD~\cite{O2 class 3}. The net baryon number susceptibility is the derivative
of free energy density with respective to the chemical potential $\mu_q$. From Eqs.~\eqref{QCD free energy} and \eqref{parameter}, we can get the derivatives of the free energy density with respect to $T$, to chemical potential $\hat{\mu}_q=\mu_q/T$ at $\hat{\mu}_q=0$, and to $\hat{\mu}_q\neq 0$ respectively,
\begin{equation}\label{scaling form of derivatives of free energy to t}
-\frac{\partial f/T^4}{\partial T^n}=-\frac{A}{(t_0T_c)^n}h^{(2-\alpha-n)/\beta\delta}f_f^{(n)}(z),
\end{equation}
\begin{equation}\label{scaling form of derivatives of free energy to mu at mu=0}
\left. -\frac{\partial f/T^4}{\partial\hat{\mu}_q^n}\right|_{\mu_q=0}=-A(2\kappa_q)^{n/2}
h^{(2-\alpha-n/2)/\beta\delta} f_f^{(n/2)}(z),
\end{equation}
\begin{equation}\label{scaling form of derivatives of free energy to mu at mu!=0}
\left. -\frac{\partial f/T^4}{\partial\hat{\mu}_q^n}\right|_{\mu_q\neq 0}=-A(2\kappa_q)^{n}(\frac{\mu_q}{T})^n
h^{(2-\alpha-n)/\beta\delta} f_f^{(n)}(z),
\end{equation}
where $n$ is even in Eq.~\eqref{scaling form of derivatives of free energy to mu at mu=0}.

Comparing Eqs.~\eqref{scaling form of derivatives of free energy to mu at mu=0} and \eqref{scaling form of derivatives of free energy to mu at mu!=0} with Eq.~\eqref{scaling form of derivatives of free energy to t}, we can see that the scaling form of derivatives of the free energy density with respect to $T$ and $\mu_q$ are equivalent. So the net baryon number fluctuations are related to the derivatives of the free energy density with respect to $T$ in the 3-dimensional $O(4)$ spin model. The $n$-th order cumulant of the energy from the 3-dimensional $O(4)$ spin model is relevant to the $2n$-th (or $n$-th ) order cumulant of the net baryon number at $\mu_q=0$ (or $\mu_q\neq0$ ) in the chiral phase transition.

The cumulants of the order parameter from the 3-dimensional $O(2)$ and $O(4)$ spin models are the derivatives of the free energy density with respect to the external field. Their critical behavior has been presented and discussed in Ref.~\cite{PanX-cpod11}.

\section{Critical behavior of the high cumulants }

The Monte Carlo simulations of the 3-dimensional Ising, $O(2)$, and $O(4)$ spin models are performed by the Wolff  algorithm with helical boundary conditions~\cite{wolff}. We start simulation without magnetic field, and then modify the algorithm to include a magnetic field $H=0.05$ and $H=0.1$~\cite{wolff-h}.

As we know, the valid region of system size of the finite-size scaling varies with magnetic field and observable. The typical size of an observable at a given magnetic field is determined by the saturation of size dependence, as shown in Ref.~\cite{Mark York}. The typical system sizes for each kind of cumulants at a given magnetic field and model are listed in Table I.

\begin{table}[htbp]
\centering
\caption{\label{comparison}The typical system size for each kind of cumulant at a given external field and model}
\begin{tabular}{|l|l|l|l|l|}
\hline
\backslashbox{field(H)}{observable}&
$\kappa_n^S(O(1))$ & $\kappa_n^E(O(1))$ & $\kappa_n^E(O(2))$& $\kappa_n^E(O(4))$\\
\hline
~~~~~~~0~ &~~~~24 & ~~~~20~ &~~~~20~&~~~~20~ \\
\hline
~~~~~~0.05 &~~~~12~ & ~~~~10~ &~~~~10~&~~~~8~ \\
\hline
~~~~~~0.1 &~~~~8~ & ~~~~8~ &~~~~8~ &~~~~8~\\
\hline
\end{tabular}
\end{table}

In order to compare the basic structure of the cumulants with and without external fields, each cumulant of the energy at vanishing and non-vanishing external fields is plotted in an identical sub-figure, and rescaled to unity by its maximum or minimum (except for the first order cumulant of the energy from $O(2)$ and $O(4)$ spin models), as shown in Fig.~1, Fig.~4, and Fig.~5. For the Ising model, the cumulants of the order parameter at vanishing and non-vanishing external fields are quite different, which may be caused by the definitions as discussed in section II below Eq.~(\ref{cumulants of order parameter}). So their cumulants at non-vanishing and vanishing external fields are presented in Fig.~2 and Fig.~3, respectively.

\begin{figure}[b]
\includegraphics[width=0.45\textwidth]{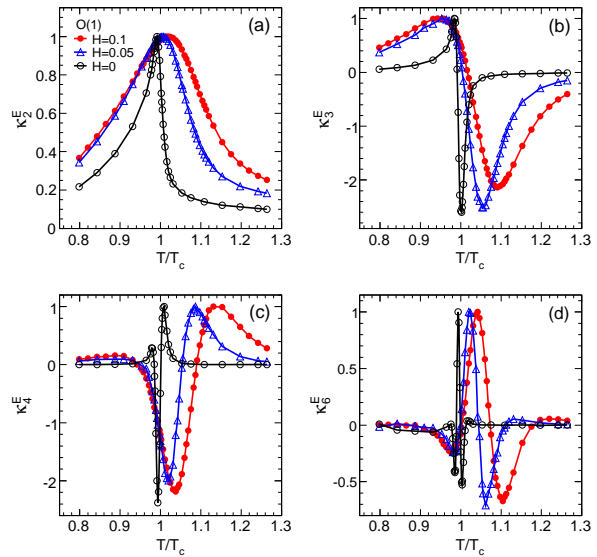}
\caption{\label{Fig. 2}(Color online). $\kappa_2^E$ (a), $\kappa_3^E$ (b), $\kappa_4^E$ (c) and $\kappa_6^E$ (d) at $H=0.1, 0.05, 0$ from the 3-dimensional Ising model.}
\end{figure}

The cumulants of the energy, i.e., $\kappa_2^E$, $\kappa_3^E$, $\kappa_4^E$ and $\kappa_6^E$, from the 3-dimensional Ising model at $H=0.1, 0.05, 0$ are shown in Fig.~1(a) to 1(d), respectively. We can see that the basic features of the cumulants, i.e., the patterns of the fluctuations, are not influenced by the magnitude of the external field. With the appearing or increasing of the external field, the whole critical region is amplified and shifted toward the high temperature side.

In the vicinity of the critical temperature, $\kappa_2^E$ has a maximum peak. $\kappa_3^E$ oscillates and its sign changes from positive to negative when temperature increases and passes the critical one. $\kappa_4^E$ has two positive maximums locating at the two sides of $T_c$, respectively. The minimum between them is negative. In contrast to the two positive maximums of $\kappa_4^E$, $\kappa_6^E$ has two negative minimums and one positive maximum at the critical temperature region. The sign change for the cumulants of the energy starts at the third cumulant. It happens twice in the fourth and sixth order cumulants.

\begin{figure}[h]
\includegraphics[width=0.45\textwidth]{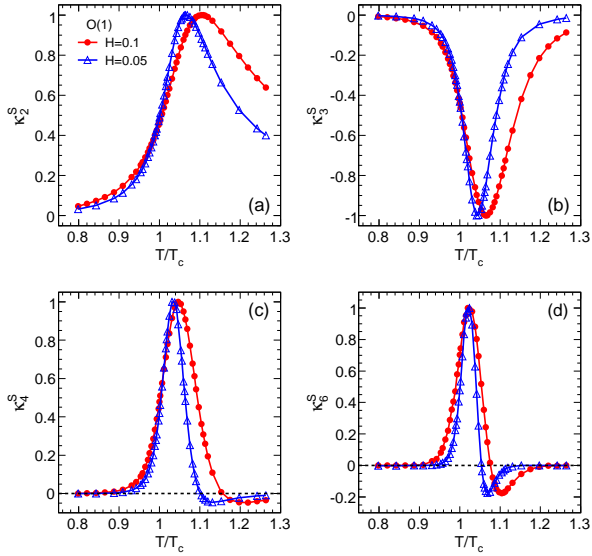}
\caption{\label{Fig. 2}(Color online). $\kappa_2^S$ (a), $\kappa_3^S$ (b), $\kappa_4^S$ (c), $\kappa_6^S$ (d) at $H=0.1, 0.05$ from the 3-dimensional Ising model.}
\end{figure}

The cumulants of the order parameter at non-vanishing external fields, i.e., $\kappa_2^S$, $\kappa_3^S$, $\kappa_4^S$ and $\kappa_6^S$, from the 3-dimensional Ising model are presented in Fig.~2(a) to 2(d), respectively. We can see that the influences of the external field are similar to those as discussed in Fig.~1.

In the vicinity of $T_c$, $\kappa_2^S$ shows the same peak structure as that for the energy. $\kappa_3^S$ has a negative valley and no sign change in the critical region. $\kappa_4^S$ shows a obvious positive maximum and a very small negative minimum when the temperature increases and passes the critical one. $\kappa_6^S$ oscillates from positive to negative, and the negative valley is more obvious than that in $\kappa_4^S$. Here the sign change starts at the fourth order cumulant in Fig.~2(c).

\begin{figure}[b]
\includegraphics[width=0.45\textwidth]{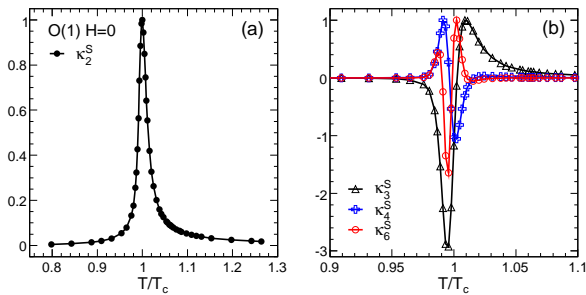}
\caption{\label{Fig. 3}(Color online). $\kappa_2^S$ (a), $\kappa_3^S$, $\kappa_4^S$ and $\kappa_6^S$ (b) at $H=0$ from the 3-dimensional Ising model.}
\end{figure}

The high cumulants of the order parameter from the 3-dimensional Ising model at vanishing external field are shown in Fig.~3(a) to 3(b), respectively. $\kappa_2^S$ in Fig.~3(a) shows a narrow and sharp peak at the critical temperature region. From Fig.~3(b), we can see that, both $\kappa_3^S$ and $\kappa_4^S$ oscillate, but the former changes from negative to positive, while the latter changes from positive to negative with the increasing temperature. The generic structure of $\kappa_6^S$ is similar to that of $\kappa_4^E$, having two positive maximum locating at the two sides of $T_c$ and a negative minimum between them. So the generic structure of the high cumulants at non-vanishing external field is quite different from that at vanishing external field. The sign change in the former case appears in the fourth order cumulant, while the third one in the later case.

Comparing the energy fluctuations in Fig.~1 with the order parameter fluctuations in Fig.~2 and Fig.~3, we can see that the generic structure of the same order cumulant of the energy is different from that of the order parameter, except for the second order one.
The way of the sign change of $\kappa_3^S$ at $H=0$ as shown in Fig.~3(b) is consistent with the expectation from the effective model~\cite{Asakawa-prl103}. The generic structure of $\kappa_4^S$ at non-vanishing external field as shown in Fig.~2(c) is very similar to that from the $\sigma$-model~\cite{Staphanov-prl107}.

\begin{figure}
\includegraphics[width=0.45\textwidth]{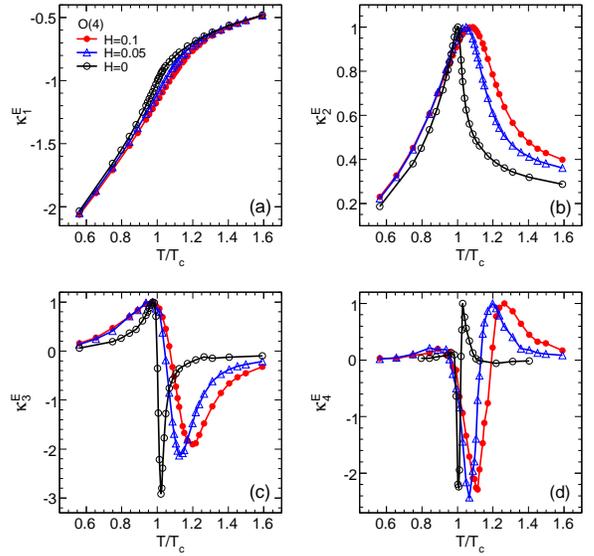}
\caption{\label{Fig. 4}(Color online). $\kappa_1^E$ (a), $\kappa_2^E$ (b), $\kappa_3^E$ (c) and $\kappa_4^E$ (d) at $H=0.1, 0.05, 0$ from the 3-dimensional $O(4)$ spin model.}
\end{figure}

The cumulants of the energy, i.e., $\kappa_1^E$, $\kappa_2^E$, $\kappa_3^E$ and $\kappa_4^E$, from the 3-dimensional $O(4)$ spin model at $H=0.1, 0.05, 0$ are presented in Fig.~4(a) to 4(d), respectively. Again, the external field shows the similar influences as discussed above.
$\kappa_1^E$ increases with the temperature. $\kappa_2^E$ has a peak and is positive in the whole critical temperature region. $\kappa_3^E$ oscillates and changes from positive to negative with the increasing temperature. $\kappa_4^E$ has two maximums and a minimum between them. The behavior of $\kappa_2^E$, $\kappa_3^E$ and $\kappa_4^E$ is similar to that of the energy fluctuations from the 3-dimensional Ising model, cf. Fig.~1(a) to 1(c).

As we discussed in section II, at vanishing baryon chemical potential, $\chi_2^B$, $\chi_4^B$, $\chi_6^B$ and $\chi_8^B$ in the chiral phase transition are corresponding to $\kappa_1^E$, $\kappa_2^E$, $\kappa_3^E$ and $\kappa_4^E$ from the 3-dimensional $O(4)$ spin model. The positive peak of $\kappa_2^E$ is consistent with $\chi_4^B$ from the calculations of lattice QCD~\cite{M. Cheng} and the estimations of the PNJL~\cite{Fu weijie-PNJL} and PQM models~\cite{PQM model}. The sign change of $\kappa_3^E$ is also observed in $\chi_6^B$ from the PQM model~\cite{six order, PQM model}.

\begin{figure}
\includegraphics[width=0.45\textwidth]{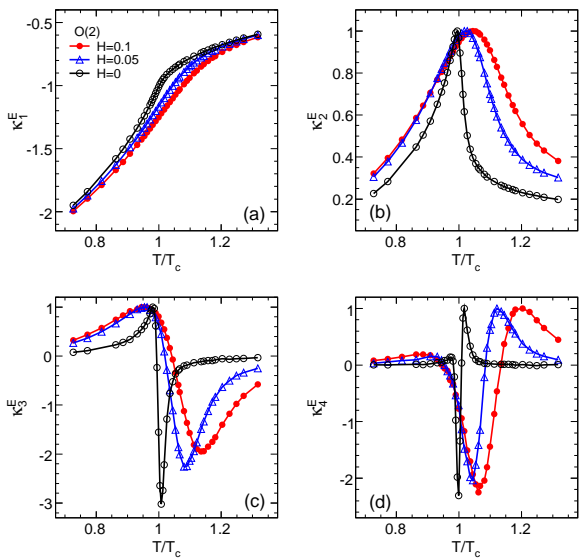}
\caption{\label{Fig. 5}(Color online). $\kappa_1^E$ (a), $\kappa_2^E$ (b), $\kappa_3^E$ (c) and $\kappa_4^E$ (d) at $H=0.1, 0.05, 0$ from the 3-dimensional $O(2)$ spin model.}
\end{figure}

The cumulants of the energy, i.e., $\kappa_1^E$, $\kappa_2^E$, $\kappa_3^E$ and $\kappa_4^E$, from the 3-dimensional $O(2)$ spin model at $H=0.1, 0.05, 0$ are presented in Fig.~5(a) to 5(d), respectively. We can see that each sub-figure in Fig.~5 is similar to that in Fig.~4. Each order cumulant of the energy from the 3-dimensional $O(2)$ and $O(4)$ spin models shows qualitatively similar temperature dependence in the vicinity of the critical temperature.

\section{Summary}

In this paper we perform the simulations of the 3-dimensional Ising, $O(2)$ and $O(4)$ spin models at a given system size at three different external fields $H=0.1, 0.05, 0$. The critical behavior of the high cumulants of the order parameter and energy in the 3-dimensional Ising model, and the cumulants of the energy in the 3-dimensional $O(2)$ and $O(4)$ spin models is presented, respectively. We find that the external field does not influence the generic structure of the cumulants, except the cumulants of the order parameter from the 3-dimensional Ising model. But it widens the temperature region of the critical fluctuations.

For the 3-dimensional Ising model, the generic structure of the high cumulants of the energy are different from that of the order parameter. For the energy fluctuations, the sign change starts at the third order
cumulant no matter with or without external field. So does the order parameter at vanishing external field.
At non-vanishing external field, the first sign change of the order parameter fluctuations appears at the fourth order cumulant.
The common feature is that the higher the order of the cumulant, the more complicated the fluctuation pattern is.

The critical behavior of the third order cumulant of the order parameter at vanishing external field, and the fourth order cumulant of the order parameter are consistent with the expectations of the effective model and the $\sigma$-model, respectively.

For the 3-dimensional $O(2)$ and $O(4)$ spin models, the behavior of the second to fourth order cumulants of the energy is similar to that from the 3-dimensional Ising model. The sign change also starts at the third order cumulant.

The net baryon number fluctuations based on the $O(4)$ spin model are qualitatively consistent with the calculations from the lattice QCD, and expectations from the PNJL and the PQM models. Our results also show that at vanishing chemical potential, the sixth order cumulant of the net baryon number is necessary in order to observe a sign change in the chiral phase transition.


The authors are grateful for valuable comments and suggestions from Prof. F. Karsch, Dr. S. Mukherjee, Dr. V. Skokov, and Dr. H. T. Ding.

This work was supported in part by the National Natural Science Foundation of China under Grant
No. 10835005 and MOE of China under Grant No. IRT0624 and B08033.

\end{document}